\documentstyle{article}

\title{Quantum Mechanics as a Classical Theory II:\\
Relativistic Theory}

\author{L.S.F. Olavo\\
Departamento de Fisica, Universidade de Brasilia,\\
70910-900 - Brasilia - D.F. - Brazil}

\begin{document}

\maketitle
\begin{abstract}
In this article, the axioms presented in the first one are reformulated
according to the special theory of relativity. Using these axioms, quantum
mechanic's relativistic equations are obtained in the presence of
electromagnetic fields for both the density function and the probability
amplitude. It is shown that, within the present theory's scope, Dirac's
second order equation should be considered the fundamental one in spite of
the first order equation. A relativistic expression is obtained for the
statistical potential. Axioms are again altered and made compatible with the
general theory of relativity. These postulates, together with the idea of
the statistical potential, allow us to obtain a general relativistic quantum
theory for {\it ensembles} composed of single particle systems.
\end{abstract}

\section{Introduction}

The first paper of this series demonstrated how, accepting a few axioms,
quantum mechanics can be derived from newtonian mechanics. Amongst these
axioms was the validity of the Wigner-Moyal Infinitesimal Transformation.
One variation of this transformation has already been amply studied\cite{%
1}-\cite{11} and the conclusion was that quantum mechanics cannot be derived
from these transformations, because the density function obtained is not
positive definite\cite{12}-\cite{15}. Various efforts to adapt a non-classical
phase space to quantum mechanics followed these frustrated attempts. These
attempts basically assume that quantum phase space cannot contain isolated
points - which would not make sense because of the uncertainty relations -
but instead regions, with dimensions related to the quantum of action. These
spaces are called stochastic phase spaces\cite{16}-\cite{23}.

We emphasize once more that the transformation here presented is distinct
from that presented in the literature cited above. In the form here
presented, it is only a mathematical instrument to obtain probability
densities in configuration space, using the limiting process already
described, from the joint probability density function defined in phase
space. In this manner it does not present the positivity problem, as was
demonstrated in the previous paper. Strictly speaking, this transformation
would not even need to constitute one of the theory's axioms, and we only
treat it as such to emphasize the differences between it and the one
generally used. It must be stressed that we are working with classical phase
space in this series of papers and that our axioms are of a purely classical
character. Moreover, we demonstrated in the first paper, that the
uncertainty relations are a consequence of the adopted formalism and not a
fundamental property of nature; so there is no reason to limit our system's
description to stochastic phase space.

In this second paper, we will show that both Klein-Gordon's and Dirac's
relativistic equations for the density function and for the probability
amplitude can be derived from small alterations in our axioms, made to adapt
quantum theory to the special theory of relativity. We will also include the
electromagnetic field in our considerations in order to obtain Dirac's
equation. Contrary to what is usually accepted, the fundamental character of
Dirac's second order equation will be established, instead of his first
order equation.

Once again, changing the postulates in order to adapt them to the general
theory of relativity and using the statistical potential concept, we will
demonstrate that it is possible to obtain a system of
general-relativistic-quantum equations, which takes into account the
gravitational field's effects, for an {\it ensemble} of one particle systems.

In the second section, we will develop the special relativistic formalism,
obtaining Klein-Gordon's and Dirac's equations for both the density function
and the probability amplitude. We will obtain an expression for the
relativistic statistical potential to be used in the general relativistic
treatment.

In the third section, we will obtain the system of general relativistic
quantum equations which includes, in the quantum mechanical treatment of one
particle system {\it ensembles}, the effects of the gravitational field.

Our conclusions will be developed in the final section.

In the appendix, we will show the relation between the density function
calculated in four dimensional space, which is a $\tau $-constant of motion,
and the density function calculated in three dimensional space, which is $t$%
-constant of motion, and also interpret the meaning of this relation.

\section{Special Relativistic Quantum Mechanics}

The {\it ensemble's} state is described by the functions $F\left( x^\alpha
,p^\alpha \right) $ where $x^\alpha $ and $p^\alpha $ are the position and
momentum four-vectors of each particle belonging to a system of the {\it %
ensemble}.

Let us list the modified axioms of our theory

\begin{description}
\item[(A1')]  Special relativistic mechanics of particles is valid for all
particles of the {\it ensemble's} component systems.

\item[(A2')]  For isolated system {\it ensembles}, the joint probability
density function is a $\tau $-constant of motion
\begin{equation}
\label{(1)}\frac d{d\tau }F\left( x^\alpha ,p^\alpha \right) =0,
\end{equation}
where $\tau $ is the proper time.

\item[(A3')]  The Wigner-Moyal Infinitesimal Transformation, defined as%
$$
\rho \left( x^\alpha +\frac{\delta x^\alpha }2,x^\alpha -\frac{\delta
x^\alpha }2\right) =\int F\left( x^\alpha ,p^\alpha \right) \exp \left( i
\frac{p^\beta \delta x_\beta }2\right) \cdot
$$
\begin{equation}
\label{(2)}\cdot \exp \left[ \frac{ie}{\hbar c}\int_0^{x+\frac{\delta x}%
2}A^\lambda \left( u\right) du_\lambda +\frac{ie}{\hbar c}\int_0^{x-\frac{%
\delta x}2}A^\lambda \left( u\right) du_\lambda \right] d^4p,
\end{equation}
where we include, for generality, an electromagnetic field through the
four-vector
\begin{equation}
\label{(3)}A^\lambda =\left( \phi ,{\bf A}\right) ,
\end{equation}
where $\phi $ is the scalar potential and ${\bf A}$ the vector potential, is
adequate for the description of a general quantum system in the presence of
electromagnetic fields.
\end{description}

With equation (\ref{(1)}), we can write
\begin{equation}
\label{(4)}\frac{dx^\alpha }{d\tau }\frac{\partial F}{\partial x^\alpha }+
\frac{dp^\alpha }{d\tau }\frac{\partial F}{\partial p^\alpha }=0.
\end{equation}
We can also use axiom (A1') to write the particle's relativistic equations
\begin{equation}
\label{(5)}\frac{dx^\alpha }{d\tau }=\frac{p^\alpha }m\quad ;\quad \frac{%
dp^\alpha }{d\tau }=f^\alpha =-\frac{\partial V}{\partial x_\alpha }.
\end{equation}

Using the transformation (\ref{(2)}) in (\ref{(4)}), we reach the expression%
$$
\frac 1{2m}\left\{ \left[ i\hbar \frac \partial {\partial y^\alpha }+\frac
ecA_\alpha \left( y\right) \right] ^2-\left[ i\hbar \frac \partial {\partial
y^{\prime \alpha }}+\frac ecA_\alpha \left( y^{\prime }\right) \right]
^2\right\} \rho -
$$
\begin{equation}
\label{(6)}-\left[ V\left( y\right) -V\left( y^{\prime }\right) \right] \rho
=0,
\end{equation}
where we once again use
\begin{equation}
\label{(7)}\frac{\partial V}{\partial x_\alpha }\delta x_\alpha =V\left( x+
\frac{\delta x}2\right) -V\left( x-\frac{\delta x}2\right) ,
\end{equation}
along with the following change of variables
\begin{equation}
\label{(8)}y^\alpha =x^\alpha +\frac{\delta x^\alpha }2\quad ;\quad
y^{\prime \alpha }=x^\alpha -\frac{\delta x^\alpha }2.
\end{equation}

If we ignore the potential term, equation (\ref{(6)}) is the Klein-Gordon's
density function equation for a spinless particle in the presence of an
electromagnetic field. If we are dealing with particles capable of coupling
to external electric and magnetic fields through their electric and magnetic
moments, $\overrightarrow{\pi }$ and $\overrightarrow{\mu }$ respectively,
then the interaction force which is a Lorentz scalar, is given, in a first
approximation, by
\begin{equation}
\label{(9)}F_{int}^\alpha =-\partial ^\alpha \left( \overrightarrow{\pi }%
\cdot {\bf E}+\overrightarrow{\mu }\cdot {\bf B}\right) .
\end{equation}

Equation (\ref{(6)}) becomes%
$$
\frac 1{2m}\left\{ \left[ i\hbar \frac \partial {\partial y^\alpha }+\frac
ecA_\alpha \left( y\right) \right] ^2-\left[ i\hbar \frac \partial {\partial
y^{\prime \alpha }}+\frac ecA_\alpha \left( y^{\prime }\right) \right]
^2\right\} \rho -
$$
\begin{equation}
\label{(10)}-\left[ \left( \overrightarrow{\pi }\cdot {\bf E}+
\overrightarrow{\mu }\cdot {\bf B}\right) \left( y\right) -\left(
\overrightarrow{\pi }\cdot {\bf E}+\overrightarrow{\mu }\cdot {\bf B}\right)
\left( y^{\prime }\right) \right] \rho =0,
\end{equation}
which we call Dirac's First Equation for the density function.

The imposition that the potential in (\ref{(10)}) be a Lorentz scalar is
enough for us to construct a tensor associated to the internal degrees of
freedom - internal moments. Land\'e's factor, cited in the last paper, can
be obtained, as usual, passing to the non-relativistic limit\cite{24} of
equation (\ref{(10)}) above.

In order to obtain an equation for the probability amplitude we can write,
in a way similar to that done in the first paper (hereafter identified as
(I)),
\begin{equation}
\label{(11)}\rho \left( y^\alpha ,y^{\prime \alpha }\right) =\Psi ^{*}\left(
y^{^{\prime }\alpha }\right) \Psi \left( y^\alpha \right) ,
\end{equation}
where
\begin{equation}
\label{(12)}\Psi \left( y^\alpha \right) =R\left( y^\alpha \right) \exp
\left[ \frac i\hbar S\left( y^\alpha \right) \right] ,
\end{equation}
being $R\left( y\right) $ and $S\left( y\right) $ real functions. Using the
change in variables (\ref{(8)}) and expanding expression (\ref{(11)}) up to
the second order in $\delta x$, we obtain%
$$
\rho \left( x^\alpha +\frac{\delta x^\alpha }2,x^\alpha -\frac{\delta
x^\alpha }2\right) =\exp \left[ \frac i\hbar \frac{\partial S}{\partial
x^\beta }\delta x^\beta \right] \cdot
$$
\begin{equation}
\label{(13)}\cdot \left\{ R\left( x^\alpha \right) ^2-\left( \frac{\delta
x^\beta }2\right) ^2\left[ \left( \frac{\partial R}{\partial x^\alpha }%
\right) ^2-R\frac{\partial ^2R}{\partial x_\beta \partial x^\beta }\right]
\right\} .
\end{equation}

Substituting this expression in equation (\ref{(6)}), written in terms of $x$
and $\delta x$ and without including, for the sake of simplicity, the
electromagnetic potentials, we get
\begin{equation}
\label{(14)}\frac{-\hbar ^2}m\frac{\partial ^2\rho }{\partial x^\alpha
\partial \left( \delta x_\alpha \right) }-\frac{\partial V}{\partial
x^\alpha }\delta x^\alpha \rho =0
\end{equation}
and, holding the zero and first order terms in $\delta x$, we reach the
equation
\begin{equation}
\label{(15)}\frac i\hbar \partial _\alpha \left( R^2\frac{\partial ^\alpha S}%
m\right) +\delta x^\alpha \partial _\alpha \left\{ \frac{-\hbar ^2}{2mR}\Box
R+V+\frac{\partial _\beta S\partial ^\beta S}{2m}\right\} =0,
\end{equation}
where we use $\partial _\alpha =\partial /\partial x^\alpha $ and $\Box
=\partial _\alpha \partial ^\alpha $. Collecting the real and complex terms
and equating them to zero, we get the pair of equations
\begin{equation}
\label{(16)}\partial _\alpha \left( R^2\frac{\partial ^\alpha S}m\right) =0,
\end{equation}
\begin{equation}
\label{(17)}\frac{-\hbar ^2}{2mR}\Box R+V+\frac{\partial _\beta S\partial
^\beta S}{2m}=const.
\end{equation}

The constant in (\ref{(17)}) can be obtained using a relativistic solution
for the free particle. In this case, it is easy to demonstrate that the
constant will be given by
\begin{equation}
\label{(18)}const.=\frac{mc^2}2,
\end{equation}
so that equation (\ref{(17)}) becomes\cite{25,26}
\begin{equation}
\label{(19)}\frac{-\hbar ^2}{2mR}\Box R+V-\frac{mc^2}2+\frac{\partial _\beta
S\partial ^\beta S}{2m}=0.
\end{equation}

Reintroducing the electromagnetic potentials, this equation is formally
identical to the equation
\begin{equation}
\label{(20)}\left\{ \frac 1{2m}\left[ i\hbar \frac \partial {\partial
x^\alpha }+\frac ecA_\alpha \left( x\right) \right] ^2+V\left( x\right)
+\left( \overrightarrow{\pi }\cdot {\bf E}+\overrightarrow{\mu }\cdot {\bf B}%
\right) \left( x\right) -\frac{mc^2}2\right\} \Psi \left( x\right) =0,
\end{equation}
since the substitution of expression (\ref{(12)}) in the equation above
gives us equation (\ref{(19)}), when the electromagnetic potentials are
considered. We call this equation, without the potential term,
Klein-Gordon's Second Equation, while for a potential such as in (\ref{(9)}%
), we call it Dirac's Second Equation for the probability amplitude.

In order to obtain mean values in relativistic phase space of some function $%
\Theta \left( x,p\right) $, we should calculate the integral
\begin{equation}
\label{(21)}\overline{\Theta \left( x^\alpha ,p^\alpha \right) }=\lim
_{\delta x\rightarrow 0}\int O_p\left( x^\alpha ,\delta x^\alpha \right)
\rho \left( x^\alpha +\frac{\delta x^\alpha }2,x^\alpha -\frac{\delta
x^\alpha }2\right) d^4x.
\end{equation}

Following the same steps as in (I), we can introduce the four-momentum and
four-position operators as being
\begin{equation}
\label{(22)}\stackrel{\wedge }{p}_\alpha ^{\prime }=-i\hbar \frac \partial
{\partial \left( \delta x^\alpha \right) }\quad ;\quad \stackrel{\wedge }{x}%
_\alpha ^{\prime }=x_\alpha .
\end{equation}

Note that we are calculating integrals in relativistic four-spaces, for it
is in these spaces that the density function is $\tau $-conserved. If we
desire to calculate values in habitual three dimensional configuration space
through the probability amplitudes we may, as is habitual, take equation (%
\ref{(20)}) for $\psi $, multiply it to the left by $\psi ^{*}$ and subtract
it from the equation for $\psi ^{*}$, multiplied for the left by $\psi $, to
obtain (in the absence of electromagnetic fields)
\begin{equation}
\label{(23)}j^\alpha =\frac{i\hbar }{2m}\left[ \psi ^{*}\partial ^\alpha
\psi -\psi \partial ^\alpha \psi ^{*}\right] ,
\end{equation}
which we define as the four-current. In this case, we have the continuity
equation
\begin{equation}
\label{(24)}\partial _\alpha j^\alpha =0,
\end{equation}
formally equivalent to equation (\ref{(16)}) if we use the decomposition (%
\ref{(12)}) for the amplitudes. In the appendix we present a different
technique to obtain this current through the density function which provides
us with the correct interpretation of $j^\alpha \left( x\right) $. We can
than take the zero component of the four-current
\begin{equation}
\label{(25)}P\left( x\right) =\frac{i\hbar }{2m}\left[ \psi ^{*}\partial
^0\psi -\psi \partial ^0\psi ^{*}\right]
\end{equation}
as being the probability density in three dimensional space, since it
reduces itself to the correct non-relativistic probability density in the
appropriate limit\cite{27}. The results referring to the existence of
particles and anti-particles are the usual and will be shortly discussed in
the appendix together with the positivity of equation (\ref{(25)}).

{}From equation (\ref{(16)}) and (\ref{(17)}), we can calculate a statistical
potential, analogous to the one obtained in the previous article. In this
case it is easy to show that this potential is given by
\begin{equation}
\label{(26)}V_{eff}\left( x\right) =V\left( x\right) -\frac{\hbar ^2}{2mR}%
\Box R,
\end{equation}
and is associated to the equation
\begin{equation}
\label{(27)}\frac{dp^\alpha }{d\tau }=-\partial ^\alpha V_{eff}\left(
x\right) ,
\end{equation}
together with the initial condition
\begin{equation}
\label{(28)}p^\alpha =\partial ^\alpha S.
\end{equation}

These last three expressions will be very useful in the next section where
we will undertake the general relativistic treatment.

\section{General Relativistic Quantum Mechanics}

The axioms should again be altered in order to make them adequate for the
general theory of relativity. Let us list our axioms bellow:

\begin{description}
\item[(A1")]  The general relativistic mechanics of particles is valid for
all particles of the {\it ensemble's} component systems.

\item[(A2")]  For an {\it ensemble} of single particle isolated systems in
the presence of a gravitational field, the joint probability density
function representing this {\it ensemble} is a conserved quantity when its
variation is taken along the system's geodesics, that is
\begin{equation}
\label{(29)}\frac{DF\left( x^\alpha ,p^\alpha \right) }{D\tau }=0,
\end{equation}
where $\tau $ is the proper time associated to the geodesic and $D/D\tau $
is the derivative taken along the geodesic defined by $\tau $.

\item[(A3")]  The Wigner-Moyal Infinitesimal Transformation defined as
\begin{equation}
\label{(30)}\rho \left( x^\alpha +\frac{\delta x^\alpha }2,x^\alpha -\frac{%
\delta x^\alpha }2\right) =\int F\left( x^\alpha ,p^\alpha \right) \exp
\left( \frac i\hbar p^\beta \delta x_\beta \right) d^4p
\end{equation}
is valid for the description of any quantum system in the presence of
gravitational fields.
\end{description}

With equation (\ref{(29)}), we can write
\begin{equation}
\label{(31)}\frac{Dx^\alpha }{D\tau }\nabla _{x^\alpha }F+\frac{Dp^\alpha }{%
D\tau }\nabla _{p^\alpha }F=0,
\end{equation}
and using axiom (A1''), we have
\begin{equation}
\label{(32)}\frac{Dx^\alpha }{D\tau }=\frac{p^\alpha }m\quad ;\quad \frac{%
Dp^\alpha }{D\tau }=f^\alpha .
\end{equation}

Using now the transformation (\ref{(30)}) and the usual change of variables
\begin{equation}
\label{(33)}y^\alpha =x^\alpha +\frac{\delta x^\alpha }2\quad ;\quad
y^{\prime \alpha }=x^\alpha -\frac{\delta x^\alpha }2,
\end{equation}
we reach the generalized relativistic quantum equation for the density
function
\begin{equation}
\label{(34)}\left\{ \frac{-\hbar ^2}{2m}\left[ \nabla _\alpha ^2-\nabla
_{\alpha ^{\prime }}^2\right] +\left[ V\left( y\right) -V\left( y^{\prime
}\right) \right] \right\} \rho =0,
\end{equation}
where $\nabla _\alpha $ and $\nabla _\alpha ^{\prime }$ are the covariant
derivatives according to $y$ and $y^{\prime }$ respectively.

Assuming the validity of the decomposition
\begin{equation}
\label{(35)}\rho \left( y^{\prime },y\right) =\Psi ^{*}\left( y^{\prime
}\right) \Psi \left( y\right) ,
\end{equation}
with
\begin{equation}
\label{(36)}\Psi \left( y\right) =R\left( y\right) \exp \left[ iS\left(
y\right) /\hbar \right] ,
\end{equation}
we obtain the following pair of expressions
\begin{equation}
\label{(37)}\nabla _\mu \left[ R\left( x\right) ^2\frac{\nabla ^\mu S}%
m\right] =0,
\end{equation}
\begin{equation}
\label{(38)}\frac{-\hbar ^2}{2mR}\Box R+V-\frac{mc^2}2+\frac{\nabla _\beta
S\nabla ^\beta S}{2m}=0,
\end{equation}
where now $\Box =\nabla ^\mu \nabla _\mu $.

Obtaining, as in the previous section, the expression for the potential and
the probability force associated with the ''statistical field''
\begin{equation}
\label{(39)}V_{\left( Q\right) }\left( x\right) =\frac{-\hbar ^2}{2mR}\Box
R\quad ;\quad f_{\left( Q\right) }^\mu =\nabla ^\mu V_{\left( Q\right)
}\left( x\right)
\end{equation}
we can write
\begin{equation}
\label{(40)}m\frac{D^2x^\mu }{D\tau ^2}=f^\mu \left( x\right) +f_{\left(
Q\right) }^\mu \left( x\right)
\end{equation}
which is an equation for the {\it ensemble}. In this case, equation (\ref
{(40)}) can be considered the equation for the possible geodesics associated
to the different configurations which the {\it ensemble's} systems may
possess. Nevertheless, we must stress that each system's particle still
obeys equation (\ref{(32)}) strictly.

According to Einstein's intuition, we put
\begin{equation}
\label{(41)}G_{\mu \nu }=-\frac{8\pi G}{c^2}\left[ T_{\left( M\right) \mu
\nu }+T_{\left( Q\right) \mu \nu }\right] ,
\end{equation}
where $G_{\mu \nu }$ is Einstein's tensor, $T_{\left( M\right) \mu \nu }$ is
the energy-momentum tensor associated to the forces represented by $f^\mu
\left( x\right) $ in equation (\ref{(40)}) and $T_{\left( Q\right) \mu \nu }$
is the tensor associated to the statistical potential.

The tensor $T_{\left( Q\right) \mu \nu }$ can be obtained looking at
equations (\ref{(37)}) and (\ref{(38)}). Equation (\ref{(38)}) represents
the possible geodesics related to the {\it ensemble}, as was pointed out
above, and equation (\ref{(37)}) defines an equation for the ''statistical''
field variables $R\left( x\right) $ and $S\left( x\right) $. The tensor
associated with this equation is given by
\begin{equation}
\label{(42)}T_{\left( Q\right) \mu \nu }=mR\left( x\right) ^2\left[ \frac{%
\nabla _\mu S}m\frac{\nabla _\nu S}m\right]
\end{equation}
and is called a matter tensor if we make the following substitution
\begin{equation}
\label{(43)}u_\mu =\frac{\nabla _\mu S}m\quad ;\quad \rho ^{\prime }\left(
x\right) =mR\left( x\right) ^2,
\end{equation}
as some kind of statistical four-velocity and statistical matter
distribution respectively, to get\cite{28}
\begin{equation}
\label{(44)}T_{\left( Q\right) \mu \nu }=\rho ^{\prime }\left( x\right)
u_\mu u_\nu
\end{equation}

The interpretation of this tensor is quite simple and natural. It represents
the statistical distribution of matter in space-time.

The system of equations to be solved is
\begin{equation}
\label{(45)}\frac{-\hbar ^2}{2mR}\Box R+V-\frac{mc^2}2+\frac{\nabla _\beta
S\nabla ^\beta S}{2m}=0,
\end{equation}
\begin{equation}
\label{(46)}G_{\mu \nu }=-\frac{8\pi G}{c^2}\left[ T_{\left( M\right) \mu
\nu }+T_{\left( Q\right) \mu \nu }\right] ,
\end{equation}
This system must be solved in the following way. First we solve Einstein's
equation for the, yet unknown, probability density $\rho ^{\prime }\left(
x\right) $ and obtain the metric in terms of this function. With the metric
at hand, expressed in terms of the functions $R\left( x\right) $ and $%
S\left( x\right) $, we return to equation (\ref{(45)}) and solve it for
these functions. The Schwartzschild general relativistic quantum mechanical
problem was already solved using this system of equations and the results
will be published elsewhere.

One important thing to note is that system (\ref{(45)},\ref{(46)}) is highly
non-linear and in general will not present quantization or superposition
effects. Also, when the quantum mechanical system is solved, we will have a
metric at hand that reflects the probabilistic character of the
calculations. This metric {\it does not} represent the real metric of
space-time; it represents a statistical behavior of space-time geometry as
related to the initial conditions imposed on the component systems of the
considered {\it ensemble}.

We will return to this matters in the last paper when epistemological
considerations will take place.

\section{Conclusion}

In this paper we derived all of special relativistic quantum mechanics for
single particle systems. Once again, we note that it is Dirac's second order
equation that is obtained; beyond this, there is nothing in the formalism
which allows us to obtain the first order equation from the second order one
through a projection operation, as is usually done. We also note that the
solutions of the linear equation are also solutions of the second order
equation, but the inverse is not true. We are thus forced by the formalism
to view Dirac's second order equation as the fundamental one, contrary to
what is accepted in the literature. It is interesting to note that there was
never any reason, other than historical, to accept Dirac's linear equation
as fundamental for relativistic quantum mechanics. This means also that we
do not need to accept as real the interpretation of the vacuum as an
antiparticle sea, since there is no need for such an entity when second
order equations are considered.

It was also possible to obtain a quantum mechanical general relativistic
equation which takes the action of gravitational fields into account. The
equations obtained pointed in the direction of quantization extinction by
strong gravitational fields.

The next paper will discuss the epistemological implications of these
results.

\appendix

\section{Three-Dimensional Probability Densities}

We have seen that we can define momentum-energy and space-time operators as
\begin{equation}
\label{(47)}\stackrel{\wedge }{p}_\alpha ^{\prime }=-i\hbar \frac \partial
{\partial \left( \delta x^\alpha \right) }\quad ;\quad \stackrel{\wedge }{x}%
_\alpha ^{\prime }=x_\alpha ,
\end{equation}
acting upon the density function. For these operators any function's mean
values are calculated using integrals defined on the volume element $d^4x$.
In relativistic quantum mechanic's usual treatment, the probability density
is defined by the expression
\begin{equation}
\label{(48)}P\left( x\right) =j^0\left( x\right) =\frac{i\hbar }{2m}\left[
\psi ^{*}\partial ^0\psi -\psi \partial ^0\psi ^{*}\right] ,
\end{equation}
as already defined in (\ref{(25)}). We now want to interpret this result and
find a connection between the density function $\rho \left( y^{\prime
},y\right) $, which is $\tau $-conserved, and the zero component of the
four-current $P\left( x\right) $, which is $t$-conserved.

To do this we start noting that the expression for the mean four-momentum is
given by
\begin{equation}
\label{(49)}\overline{p^\alpha }=\lim _{\delta x\rightarrow 0}\int -i\hbar
\frac \partial {\partial \left( \delta x_\alpha \right) }\rho \left(
x^\alpha +\frac{\delta x^\alpha }2,x^\alpha -\frac{\delta x^\alpha }2\right)
d^4x.
\end{equation}

Supposing that we can decompose the density function according to
\begin{equation}
\label{(50)}\rho \left( x^\alpha +\frac{\delta x^\alpha }2,x^\alpha -\frac{%
\delta x^\alpha }2\right) =\Psi ^{*}\left( x^\alpha +\frac{\delta x^\alpha }%
2\right) \Psi \left( x^\alpha -\frac{\delta x^\alpha }2\right)
\end{equation}
and substituting it in expression (\ref{(49)}), it can be shown that
\begin{equation}
\label{(51)}\overline{p^\alpha }=\int \frac \hbar {2i}\left[ \Psi \left(
x\right) \partial ^\alpha \Psi ^{*}\left( x\right) -\Psi ^{*}\left( x\right)
\partial ^\alpha \Psi \left( x\right) \right] d^4x,
\end{equation}
or, in terms of the four-current
\begin{equation}
\label{(52)}\overline{p^\alpha }=\int mcj^\alpha \left( x\right) d^4x
\end{equation}

Thus, for a closed system, we can guarantee that the integral of the zero
component of the above vector, the energy, will not vary in time. In fact,
since
\begin{equation}
\label{(53)}\partial _\alpha j^\alpha =0,
\end{equation}
we can write the above integral as an integral only in three dimensional
space\cite{29}. To obtain a dimensionless value, we can divide the
expression (\ref{(52)}) by $\pm mc$. In this manner, we guarantee that the
integral of the zero component of the four-vector in (\ref{(52)}) is a $t$%
-conserved dimensionless function. And more, it can be shown that the term $%
j^0\left( x\right) $ reduces to the probability density in the
non-relativistic limit.

Nevertheless, there is one last step towards the acceptance of $j^0\left(
x\right) $ as a probability density. This function, as is well known, can
have both positive and negative values. This is expected if we consider that
relativistic energy has this characteristic. That's why we dived by $\pm mc$
for the particle and anti-particle respectively to obtain positive definite
probabilities. With this procedure we distinguish particles and
anti-particles in the mathematical formalism without any need to multiply by
electronic charges as is usually done.

With these conventions we have obtained the probability density for three
dimensional space.

\end{document}